\documentclass[12pt,a4]{article}
\usepackage[dvips]{graphics,color}
\usepackage{graphpap}
\usepackage{latexsym}
\begin{document}
\newcommand{\beq}{\begin{equation}}
\newcommand{\eeq}{\end{equation}}
\newcommand{\beqn}{\begin{eqnarray}}
\newcommand{\eeqn}{\end{eqnarray}}
\newcommand{\dpf}{\displaystyle\frac}
\newcommand{\no}{\nonumber}
\newcommand{\ep}{\epsilon}
\begin{center}
{\large Evolving of a massless scalar field in Reissner--Nordstr\"{o}m
Anti--de Sitter spacetimes}
\end{center}
\vspace{1ex}
\centerline{\large B. Wang$^{a,b,}$\footnote[1]{e-mail:binwang@fma.if.usp.br},
\ C. Molina  $^{a,}$\footnote[2]{e-mail:cmolina@fma.if.usp.br}
and E. Abdalla$^{a,}$\footnote[3]{e-mail:eabdalla@fma.if.usp.br}
}
\begin{center}
{$^{a}$ Instituto De Fisica, Universidade De Sao Paulo,
C.P.66.318, CEP
05315-970, Sao Paulo, Brazil \\
$^{b}$ Department of Physics, Shanghai Teachers' University,
P. R. China}
\end{center}
\vspace{6ex}
\begin{abstract}
We investigate the evolution of a scalar field propagating in
Reissner--Nordstr\"{o}m Anti--de Sitter spacetime. Due to the
characteristic of spacetime geometry, the radiative tails associated with
a massless scalar field propagation have an oscillatory exponential decay.
The object--picture of the quasinormal ringing has also been obtained. For
small charges, the approach to thermal equilibrium is faster for larger
charges. However, after the black hole charge reaches a critical value, we
get the opposite behavior for the imaginary frequencies of the quasinormal
modes. Some possible explanations concerning the wiggle of the imaginary
frequencies have been given. The picture of the quasinormal modes
depending on the
multipole index has also been illustrated.
\end{abstract}
\vspace{6ex} \hspace*{0mm} PACS number(s): 04.30.Nk, 04.70.Bw
\vfill
\newpage
\section{Introduction}

The study of wave dynamics outside black holes has been an intriguing
subject for the last few decades (for a review,  see \cite{KoK-Sch}).
In virtue of
previous works, we now have the schematic picture regarding the
dynamics of waves outside a spherical collapsing object. A static
observer outside the black hole can indicate three successive stages
of the wave evolution. First the exact shape of the wave front
depends on the initial pulse. This stage is followed by a quasi--normal
ringing, which carries information about the structure of the
background spacetime and is believed to be a unique fingerprint to
directly identify the black hole existence. Detection of these
quasinormal modes is expected to be realized through gravitational
wave observation in the near future \cite{KoK-Sch}. Finally, at late
times, quasinormal oscillations are swamped by the relaxation
process. This relaxation is the requirement of the black hole no--hair
theorem \cite{Mis-Tho-Whe}. Besides the dynamical mechanism of
shedding the perturbation hair near black hole event horizon is of
direct interest to the problem of stability of Cauchy horizons
\cite{Poi-Isr}.  

The mechanism responsible for the relaxation of neutral external
perturbations was first exhibited by Price \cite{Pri}. Studying the
behavior of a massless scalar field propagating on a fixed
Schwarzschild background, he showed that for a fixed position the
field dies off with a power--law tail. The behavior of neutral
perturbations along null infinity and future event horizon was further
studied by Gundlach, Price and Pullin \cite{Gun-Pri-Pul883} and
similar power--law tails have been obtained. These results were later
confirmed using several different techniques, both analytic and
numerical \cite{Gom-Win-Sch,Lev,And97}, and were
generalized to {\it Reissner--Nordstr\"{o}m} (RN) background
\cite{Gun-Pri-Pul883,Bic}. The application of linear approaches is
encouraged by numerical analysis of the fully nonlinear dynamics of
the fields \cite{Gun-Pri-Pul890,Bur-Ori},   which indicates the same
late time pattern of decay. 

Extending the basic scenario to study the asymptotic evolution of
charged fields and self--interacting massive scalar field around a RN
black hole, Hod and Piran \cite{Hod-Pir17,Hod-Pir18} found that
although usual inverse power--law relations of these fields present at
timelike infinity and null infinity, along the future black hole event
horizon this power--law tail is accompanied by an oscillatory
behavior. Recently, Brady et
al. \cite{Bra-Cha-Laa-Poi,Bra-Cha-Kri-Lag} studied scalar wave
dynamics in  non--asymptotically flat exteriors of Schwarzschild--de
Sitter and RN de Sitter black holes. Contrary to the asymptotically
flat geometries, no power--law tails were detected in these
cases. Instead the waves were found to decay exponentially at late
times. For $l=0$, they found that the field does not decay, but
settles down to a nonzero constant. Moreover, for a field strongly coupled
to curvature, they obtained that the wave function oscillates with an
exponentially decaying amplitude. These observations support the
earlier argument by Ching et al. \cite{Chi-Leu-Sue-You} that usual
inverse power--law tails as seen in asymptotically flat black hole
spacetimes are not a general feature of wave propagation in curved
spacetime. Besides some relation to the perturbative field, the
relaxation process reflects a characteristic of the background geometry.

It is of interest to extend this study to {\it Anti--de Sitter} (AdS)
spacetime. In addition to three major aspects that the evolution of
test--field is associated with, including the no--hair theorem, the
stability of Cauchy horizon and direct evidence of the existence of
black hole provided by quasinormal ringing, the recent discovery of
the {\it Anti--de Sitter/Conformal Field Theory} (AdS/CFT) correspondence
makes the investigation in AdS black hole background more
appealing. The quasinormal frequencies of AdS black hole have direct
interpretation in terms of the dual {\it Conformal Field Theory} (CFT). In
terms of the AdS/CFT correspondence \cite{Mal,Wit,Gub-Kle-Pol}, the
black hole corresponds to an approximately thermal state in the field
theory, and the decay of the test--field corresponds to the decay of
the perturbation of the state. The first study of the scalar
quasinormal modes in AdS space was performed by Horowitz and Hubeny
\cite{Hor-Hub} on the background of Schwarzschild AdS black holes in
four, five and seven dimensions. They claimed that for large black
holes both the real and imaginary parts of the quasinormal frequencies
scale linearly with the black hole temperature. The time scale for
approaching  the thermal equilibrium is detected by the imaginary part
of the lowest quasinormal frequency and is proportional to the inverse
of the black hole temperature. Considering that the RN AdS solution
provides a better framework than the Schwarzschild AdS geometry and
may contribute significantly to our understanding of space and time,
we generalized the study made in our previous work \cite{Wan-Abd}. We
found that the charge in RN AdS black hole showed a richer physics
concerning quasinormal modes and further information on AdS/CFT
correspondence. The bigger the black hole charge is, the quicker for
their approach to thermal equilibrium in CFT. However these studies
focused much on frequencies of the quasinormal modes, the
object--picture of the evolution of test--field around the AdS
background is lacking.  

The intention of this paper is to analyse in detail the wave
propagation of massless scalar field in RN AdS spacetime. We will show
that the direct picture of the evolution presents us with a perfect
agreement
on quasinormal frequencies with those obtained by using approximation
method suggested in \cite{Hor-Hub}. Moreover, our work will show
object--pictures of the behavior of quasinormal modes as a function
of the charge $Q$ and of the multipole order of the field $l$. We will
also
address some discussions on highly charged background case. In
addition, the relaxation process at the event horizon will also be
discussed in our direct picture. We found that the decay has the
pattern of oscillatory exponential tail. This result supports
Horowitz's claim that there are no power--law tails at late times in
AdS space. Some physical explanation related to this result will be
given. 
\section{Equations and numerical methods}      

The  Reissner--Nordstr\"{o}m black hole solution of Einstein's
equations in free space with a negative cosmological constant $\Lambda
= - 3/R^2$ is given by 
\beq    
{\rm d}s^2=-h{\rm d}t^2+h^{-1}{\rm d}r^2+r^2{\rm d}\Omega^2, 
\eeq 
with 
\beq    
h=1-\dpf{r_+}{r}-\dpf{r_+^3}{R^2 r}-\dpf{Q^2}{r_+ r}+\dpf{Q^2}{r^2}+\dpf{r^2}{R^2}.  
\eeq 
The asymptotic form of this spacetime is AdS. The mass of the black hole is 
\beq          
M=\dpf{1}{2}(r_+ +\dpf{r_+^3}{R^2}+\dpf{Q^2}{r_+}).  
\eeq 
The Hawking temperature is given by the expression 
\beq  
T_H =\dpf{1-\dpf{Q^2}{r_+^2}+\dpf{3r_+^2}{R^2}}{4\pi r_+} 
\eeq 
and the potential by 
\beq 
\phi =\dpf{Q}{r_+} 
\eeq 
In the extreme case
$r_+$ and $Q$ satisfy the relation 
\beq   
1-\dpf{Q^2}{r_+^2}+\dpf{3r_+^2}{R^2}=0.  
\label{ext-case}
\eeq

For a non--extreme RN AdS black hole, the spacetime possesses two
horizons, namely the black hole event horizon $r_+$ and Cauchy horizon
$r_-$. For the extreme case where (\ref{ext-case}) is satisfied, these
two horizons degenerate. The function $h$ has four zeros at $r_+, r_-$
and $r_1, r_2$, where $r_1, r_2$ are two complex roots with no physical
meaning. The relations between these four roots are 
\beqn        
r_1+r_2 & = & -(r_++r_-)                 \no \\
r_1r_2 & = & R^2+ r_+r_- +r_+^2+r_-^2.               
\eeqn
In terms of these quantities, $h$ can be expressed as 
\beq          
h=\dpf{1}{R^2r^2}(r-r_+)(r-r_-)(r-r_1)(r-r_2).
\eeq

Introducing the surface gravity $\kappa_i$ associated with $r_i$ by
the relation 
$\kappa_i=\dpf{1}{2}\left| \dpf{dh}{dr}\right|_{r=r_i}$, we have
\beqn        
\kappa_{r_+} & = & \dpf{1}{2R^2}\dpf{(r_+-r_-)(r_+-r_1)(r_+-r_2)}{r_+^2} \no \\
\kappa_{r_-} & = & \dpf{1}{2R^2}\dpf{(r_+-r_-)(r_--r_1)(r_--r_2)}{r_-^2} \no \\
\kappa_{r_1} & = & \dpf{1}{2R^2}\dpf{(r_+-r_1)(r_--r_1)(r_1-r_2)}{r_1^2} \no \\
\kappa_{r_2} & = & \dpf{1}{2R^2}\dpf{(r_+-r_2)(r_--r_2)(r_2-r_1)}{r_2^2}
\eeqn
These quantities allow us to write
\beq     
h^{-1}=\dpf{1}{2\kappa_{r_+}(r-r_+)}-\dpf{1}{2\kappa_{r_-}(r-r_-)}+
\dpf{1}{2\kappa_{r_1}(r-r_1)}-\dpf{1}{2\kappa_{r_2}(r-r_2)}
\eeq
Combining the last two terms in (10), we express the transformation
between $r$ and the ``tortoise coordinate''  $r^*=\int h^{-1}dr$ in the form
\beqn         
r^* & = & \dpf{1}{2\kappa_{r_+}}\ln (r-r_+) -\dpf{1}{2\kappa_{r_-}}\ln (r-r_-) \no \\
    &   &
+\dpf{R^2[(r_1r_2)^2-r_+r_-r_1r_2]}{[r_+^2-r_+(r_1+r_2)+r_1r_2][r_-^2-r_-(r_1+r_2)+r_1r_2]}
\int\dpf{dr}{r^2-r(r_1+r_2)+r_1r_2} \no \\
    &   &
+\dpf{R^2[r_+r_-(r_1+r_2)-r_+r_1r_2-r_-r_1r_2]}
{[r_+^2-r_+(r_1+r_2)+r_1r_2][r_-^2-r_-(r_1+r_2)+r_1r_2]} 
\int \dpf{rdr}{r^2-r(r_1+r_2)+r_1r_2}. 
\eeqn

In terms of $t$ and $r^*$ we introduce null coordinates $u=t-r^*$ and
$v=t+r^*$ so that the future black hole horizon is located at
$u=\infty$. Since the quasinormal modes of AdS space are defined to be
modes with only ingoing waves near the horizon, we will pay more
attention on the wave dynamics near the event horizon.  

Let us consider a massless scalar field $\Phi$ in the RN AdS
spacetime, obeying the wave equation
\beq 
\Box \Phi =0
\eeq
where $\Box=g^{\alpha\beta}\nabla_{\alpha}\nabla_{\beta}$ is the d'Alembertian operator.
If we decompose the scalar field according to 
\beq    
\Phi=\sum_{lm}\dpf{1}{r}\psi _l (t,r)Y_{lm}(\theta, \phi)
\eeq
then each wave function $\psi _l (r)$ satisfies the equation
\beq       
-\dpf{\partial^2 \psi _l}{\partial t^2}+\dpf{\partial^2\psi _l}{\partial r*^2}=V_l\psi
_l,
\eeq
where
\beqn       
V_l & = & h\left[\dpf{l(l+1)}{r^2}+\dpf{1}{r}\dpf{dh}{dr}\right] \no \\
        & = & h\left[\dpf{l(l+1)}{r^2}+\dpf{r_+ +r_+
^3/R^2+Q^2/r_+}{r^3}-\dpf{2Q^2}{r^4}+\dpf{2}{R^2}\right].
\eeqn

The potential $V_l$ has the same characteristic as that in
Schwarzschild AdS black hole. It is positive and vanishes at the
horizon, but diverges at $r \rightarrow\infty$, which requires that
$\Phi$ vanishes at infinity. This is the boundary condition to be
satisfied by the wave equation for the scalar field in AdS space.  In
terms of the radial coordinate $r^*$, it is seen that when $r$ tends
to infinity, $r^*$ tends to a finite constant (which we denote
$r^*_{as}$). It means that our region of interest in the $(u-v)$
diagram is above the line $v - u = 2 r^*_{as}$, as shown in figure 1.
In this line (where $r \rightarrow \infty$) we set $\Phi = 0$.

The behavior of the potential differs quite a lot from that of asymptotically flat space
and de Sitter space. As argued in \cite{Chi-Leu-Sue-You}, it is this
peculiarity that contributes to the special wave propagation as will
be shown later. 

Using the null coordinates $u$ and $v$, the equation (14) can be recast as 
\beq  
-4\dpf{\partial^2}{\partial u\partial v}\psi_l(u,v)=V_l(r)\psi_l(u,v)
\eeq
in which $r$ is determined by inverting the relation $r^*(r)=(v-u)/2$. 

The two--dimensional wave equation (16) can be integrated numerically,
using for example the finite difference method suggested in
\cite{Gun-Pri-Pul883,Gom-Win-Isa98}. Using Taylor's theorem, it is
discretized as 
\beq 
\psi_N=\psi_E+\psi_W-\psi_S-\delta u\delta v
V_l\left(\dpf{v_N+v_W-u_N-u_E}{4}\right)\dpf{\psi_W+\psi_E}{8}
+\mathcal{O}(\epsilon^4)
\label{disc-eq}
\eeq
where the points $N, S, E$ and $W$ form a null rectangle with relative
positions as: \linebreak $N: (u+\delta u, v+\delta v), W: (u+\delta u, v), E: (u,
v+\delta v)$ and $S: (u, v)$. The parameter $\epsilon$ is an overall grid scalar
factor, so that $\delta u \sim  \delta v \sim \epsilon$.

Considering that the behavior of the wave function is not sensitive to the
choice of initial data, we set $\psi_l(u, v=v_0)=0$ and use a Gaussian
pulse as an initial perturbation, centered on $v_c$ and with
width $\sigma$ on $u=u_0$ as 
\beq       
\psi_l(u=u_0, v)=\exp\left[-\dpf{(v-v_c)^2}{2\sigma^2}\right]
\eeq

The inversion of the relation $r^*(r)$ needed in the evaluation of the
potential $V_l(r)$ is the most tedious part in the computation. We
overcome this difficulty by employing the method suggested in
\cite{Gun-Pri-Pul883,Bra-Cha-Laa-Poi}.  

\begin{center}
\setlength{\unitlength}{1.0mm}
\begin{picture}(150,75)(0,0)
\put(0,0){\resizebox{76\unitlength}{75\unitlength}%
{\includegraphics{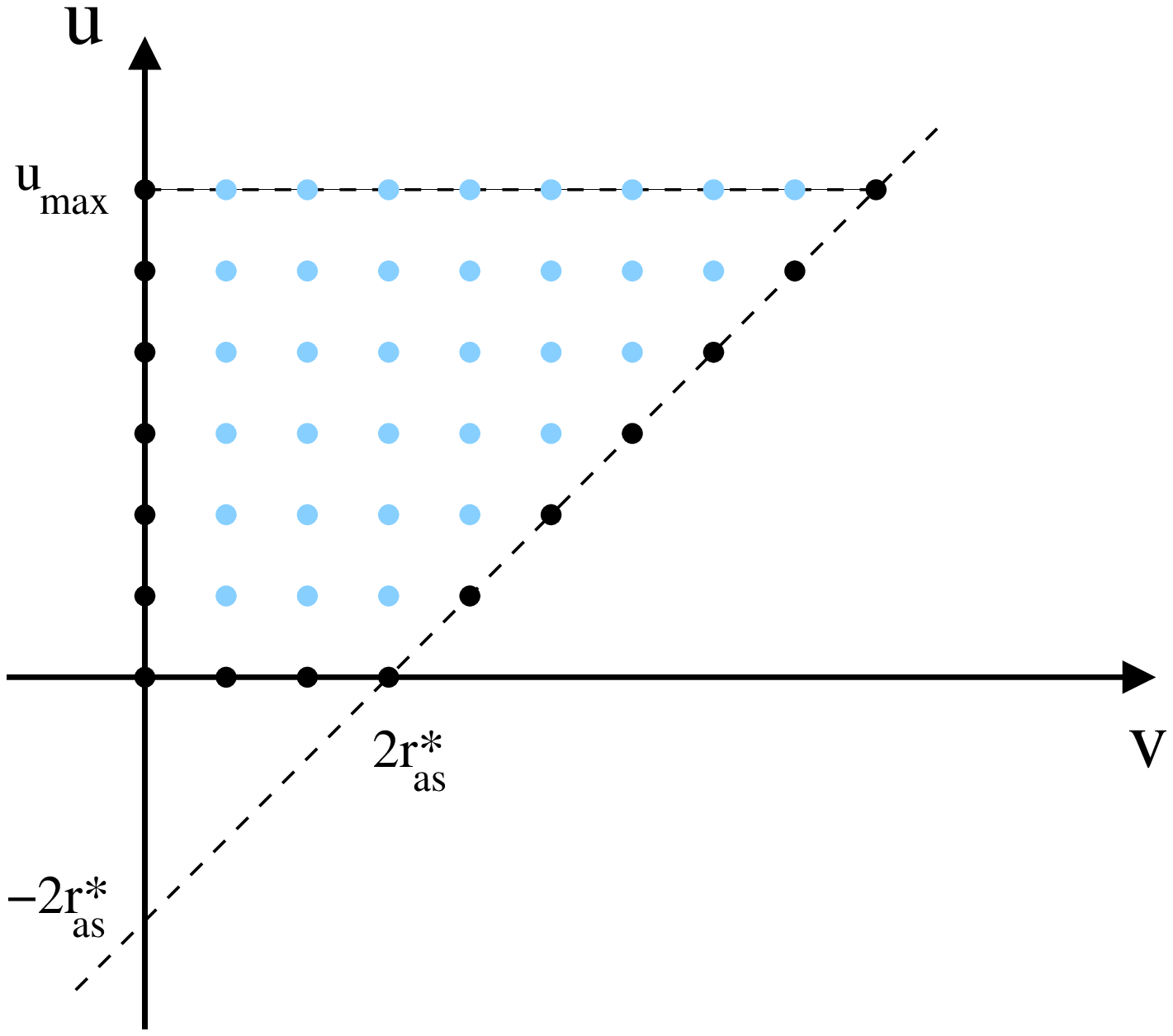}}}
\put(85,10){\resizebox{65\unitlength}{55\unitlength}%
{\includegraphics{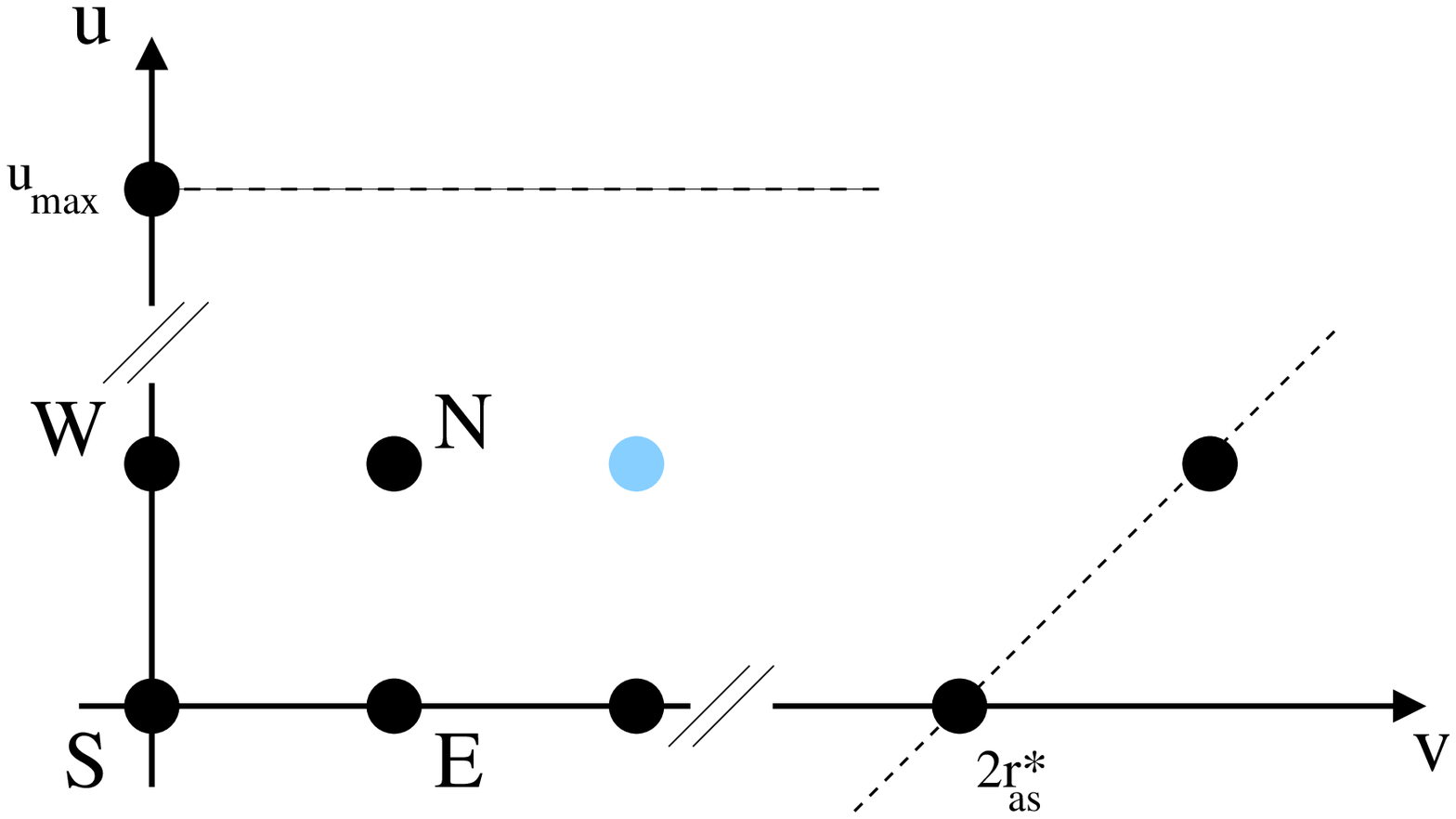}}}
\end{picture} 
 \parbox[t]{\textwidth}{\small Figure 1: {\it  (left) Diagram of the
 numerical grid and the domain of interest. The black spots represent
the grid points where the value of the field is known. The gray spots
represent the grid points to be calculated. The points in the line $u
= u_{max}$ are the results shown in this paper. (right) Detail of the
 previous diagram, showing the relative positions N, S, E and W. The
gray point would be the next one to be calculated. }}
\end{center}

After the integration is completed, the value $\psi_l(u_{max}, v)$ are
extracted,  where  $u_{max}$  is  the  maximum value  of  $u$  on  the
numerical grid. Taking  sufficiently large $u_{max}$, $\psi_l(u_{max},
v)$ represents a good approximation for the wave function at the event
horizon ($u \rightarrow \infty$),  which carries information about the
quasinormal modes  for AdS space of  our interest. It  was observed in
our numerical experiments that the plots obtained converged in a given
range of $v$, as expected, but the rate of convergence varied with the
parameters  of  the   system.  It  should  be  noted   that  the  term
``convergence'' in the last sentence  refers to the limit of $\Phi$ as
$u_{max}$ is taken bigger and bigger. As commented, the region
$u_{max} \rightarrow \infty$  corresponds to the  event horizon, in which
we are interested.  Some  comments about the convergence
of the numerical code,  meaning the evolution  of the wave function
with the grid size, is made in appendix A. In order to compare our
results here with those in \cite{Hor-Hub,Wan-Abd}, we fix $R=1$ in the
following.
\section{Numerical results}

We now report on the results of our numerical simulations of evolving
massless scalar field on a RN AdS black hole background. Taking
$Q\rightarrow 0$, our results reflect the properties on
Schwarzschild AdS background.

\subsection{Behavior of wave evolution for $l=0$}

In the first series of numerical experiments, we choose the multipole
index $l=0$ and examine the behavior of the test--field propagation
with  the increase of the charge of the background spacetime.

\begin{center}
\setlength{\unitlength}{1.0mm}
\begin{picture}(150,75)(0,0)
\put(0,0){\resizebox{76\unitlength}{75\unitlength}%
{\includegraphics{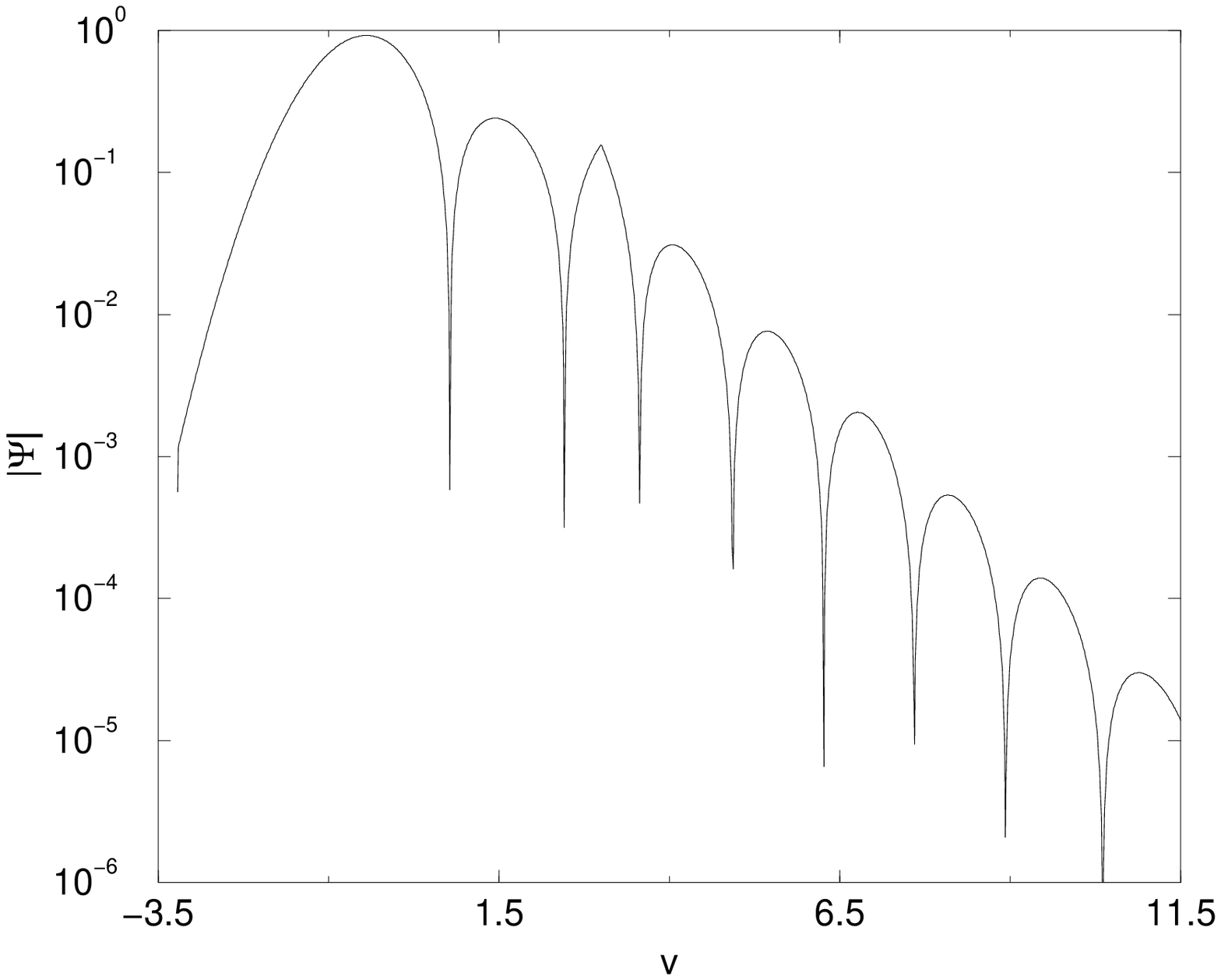}}}
\put(80,0){\resizebox{76\unitlength}{75\unitlength}%
{\includegraphics{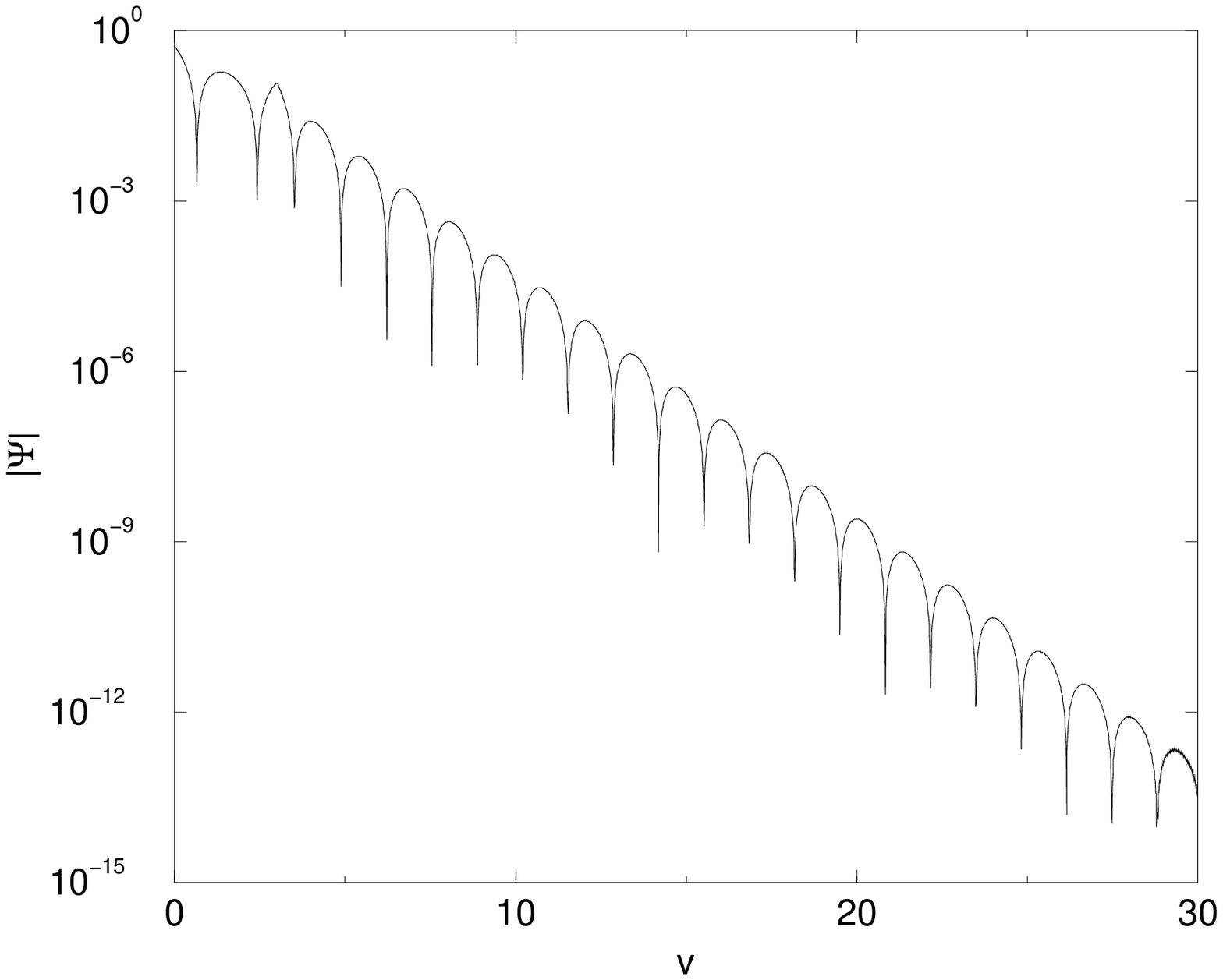}}}
\end{picture}
\parbox[t]{\textwidth}{\small Figure 2: {\it Semi--log graph of the
absolute value of the wave function for small values of $v$ (left) and
the late time behavior (right), with $r_+ = 0.4$ and $Q=0$.  }}
\end{center}

The figure 2 displays the quasinormal ringing on the 4D
Schwarzschild--AdS background. From plots using different values of
$r_+$ we can read off values for the imaginary ($\omega_I$) and real 
($\omega_R$) parts of the frequency. The results are shown in table 1.

\begin{center}
\begin{tabular}{|c||c|c|}  \hline \hline 
 $r_+$ & $\omega^1_I$ & $\omega^1_R$ \\ \hline \hline 
100 & 274.61 & 185.38  \\ \hline 
50  & 133.68 & 91.38  \\ \hline 
10  & 26.79  & 18.85 \\ \hline 
5   & 13.41  & 9.99  \\ \hline 
1   & 2.67   & 2.79  \\ \hline 
0.8 & 2. 15  & 2.58  \\ \hline 
0.6 & 1.58   &  2.41 \\ \hline 
0.4 & 1.006  & 2.362 \\ \hline  \hline
\end{tabular}

\vspace{0.5cm}

\parbox[t]{\textwidth}{\small Table 1: {\it The lowest quasinormal
mode frequency for four--dimensional Schwarzschild--AdS black hole
for $l=0$. The real part is $\omega^1_R$ and the imaginary part is
$\omega^1_I$.}}

\end{center}

The agreement with the frequencies calculated in \cite{Hor-Hub} is
good. A small difference can be attributed to the different methods
employed.

We studied the late--time decay of the test--field on
Schwarzschild--AdS black hole and the results are shown in figure
2. It was first predicted in \cite{Hor-Hub} that the decay is always
exponential and with no power--law tails. Through careful study, we
got a result that supports their claim.  We obtain that the late--time
falloff is oscillatory exponential. Recall that for the AdS black
hole, the potential diverges at infinity but vanishes exponentially
near the black hole horizon.  This boundary condition differs
completely from that of asymptotically flat spacetime. Using Ching et
al's argument \cite{Chi-Leu-Sue-You}, it is not difficult to
understand the reason that usual power--law tail is replaced by
exponentially decay.  Since that potential also falls off
exponentially on black hole horizon in de Sitter space, the pattern of
decay exhibit here on AdS black hole event horizon is similar to that
in de Sitter space \cite{Bra-Cha-Laa-Poi,Bra-Cha-Kri-Lag}.  However
due to the waves bouncing off the divergent potential barrier at large
$r$ in AdS space, the oscillation appears in the exponential tail.
With the late--time behavior in hand, it is possible to improve
previous studies on Cauchy horizon stability problems in AdS black
holes \cite{Hel-Kon,Wan-Su} along the lines of Brady and Smith
\cite{Bra-Smi}.

\begin{center}
\setlength{\unitlength}{1.0mm}
\begin{picture}(110,80)
\put(0,-2){\resizebox{110\unitlength}{80\unitlength}
{\includegraphics{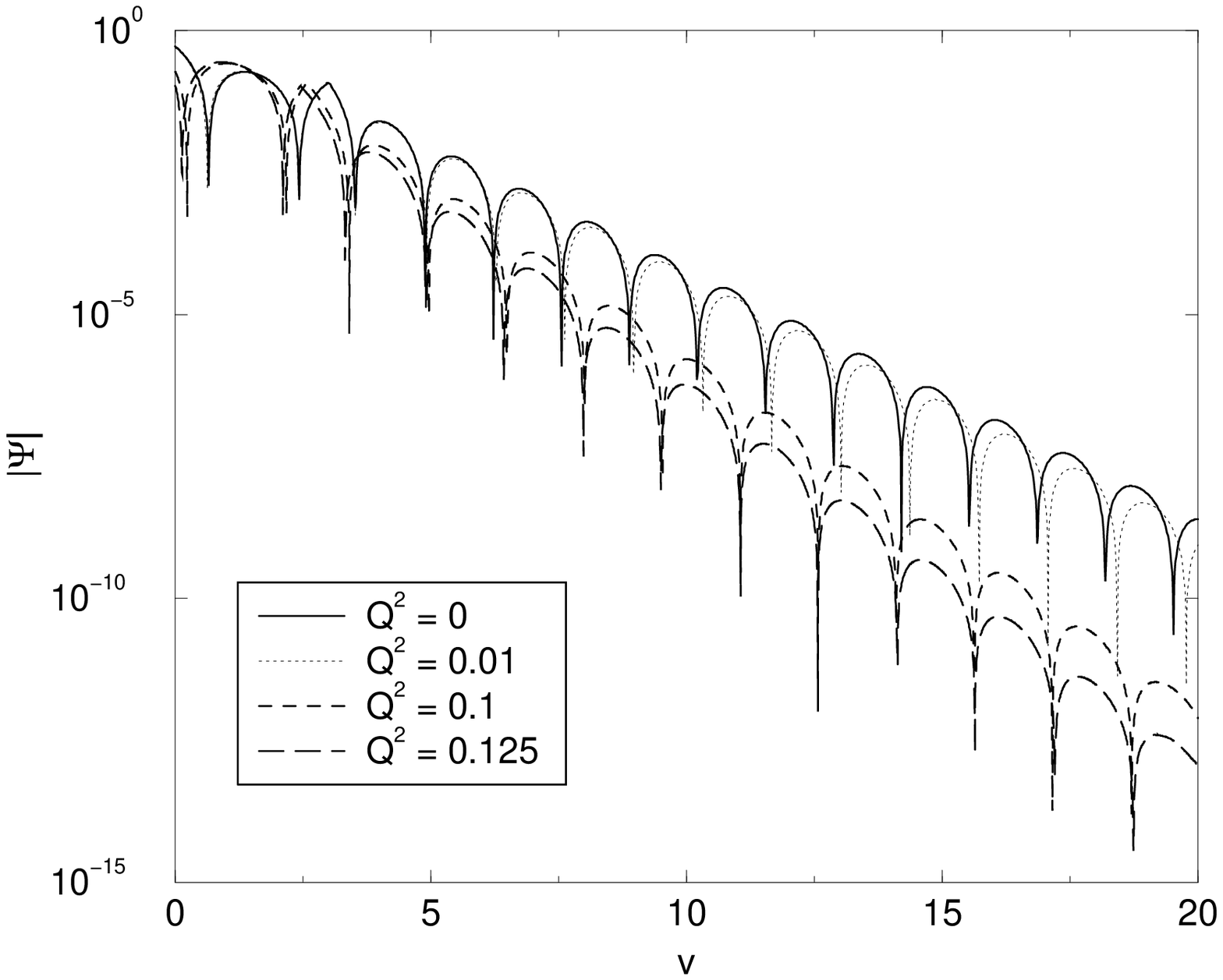}}}
\end{picture}
\parbox[t]{\textwidth}{Figure 3: {\it \small Semi--log graphs of $|\Psi|$
with $r_+ = 0.4$ and small values of $Q$. The extreme value for $Q^2$
is 0.2368.}}
\end{center}

\begin{center}
\begin{tabular}{|c|c||c|c|}  \hline \hline 
  \multicolumn{4}{|c|}{$r_+ = 0.4$ ($Q^2_{ext} = 0.2368$)} \\ \hline \hline 
$Q^2$ & $r_-$ & $\omega^1_I$ & $\omega^1_R$ \\ \hline \hline 
0     & 0      &  1.007        & 2.363         \\ \hline  
0.01  & 2.14E-2 & 1. 034 &  2.327    \\ \hline   
0.1    & 0.196 & 1.42 & 2.05     \\ \hline   
0.125  & 0.238 & 1.53 & 2.04 \\ \hline  \hline
\end{tabular}

\begin{tabular}{|c|c||c|c|}  \hline \hline 
\multicolumn{4}{|c|}{$r_+ = 1$ ($Q^2_{ext} = 4$)} \\ \hline \hline
$Q^2$ & $r_-$ & $\omega^1_I$ & $\omega^1_R$ \\ \hline \hline 
0 & 0   & 2.67 & 2.79          \\ \hline  
0.01  & 4.9875E-003 & 2.68 & 2.78 \\ \hline \hline
\end{tabular}

\vspace{0.5cm}

\parbox[t]{\textwidth}{\small Table 2: {\it The behavior of the lowest
 quasinormal mode frequency with the increase of the charge in RN AdS
 black hole background for $l=0$. The real part is $\omega^1_R$ and
 the imaginary part is $\omega^1_I$.}}
\end{center}

Figure 3 demonstrates the behaviors of the field with the increase of
the charge in RN AdS black hole background. In table 2 we listed
values of imaginary and real parts of quasinormal frequencies read
from the plots. These frequencies together with the entire picture
presented in figure 3 agree perfectly with the result given in
\cite{Wan-Abd}. Since the imaginary and real parts of the
quasinormal frequencies relate to the damping time scale
($\tau_1=1/\omega_i$) and oscillation time scale
($\tau_2=1/\omega_r$), respectively. We learned that as $Q$
increase, $\omega_i$ increase as well, which corresponds to the
decrease of the damping time scale. According to the AdS/CFT
correspondence, this means that for bigger $Q$, it is quicker for
the quasinormal ringing to settle down to  thermal
equilibrium. From figure 3 it is easy to see this
property. Besides, table 2 and figure 3 also tell us that the
bigger $Q$ is, the lower frequencies of oscillation will be. This
object--picture support the argument in \cite{Wan-Abd} that if we
perturb a RN AdS black hole with high charge, the surrounding
geometry will not ``ring'' as much and as long as that of the black
hole with small $Q$. It is easy for the perturbation on the high
charged AdS black hole background to return to thermal
equilibrium. However this relation seems  not to hold well when the
charge is sufficiently big and near the extreme value described by
relation (\ref{ext-case}). We will address this behavior later.

\subsection{Behavior of wave evolution with the increase of $l$}

We have so far discussed only of the lowest multipole index $l=0$. In
the following we show the wave dynamics on the RN AdS background for
different multipole indices $l$. For simplicity we present here the
results for $l=0,1,2$, but the basic characteristics of the
dependence of the wave function with the multipole index have already
been uncovered and are discussed in this section.

\begin{center}
\setlength{\unitlength}{0.8mm}
\begin{picture}(110,80)
\put(0,-2){\resizebox{110\unitlength}{80\unitlength}
{\includegraphics{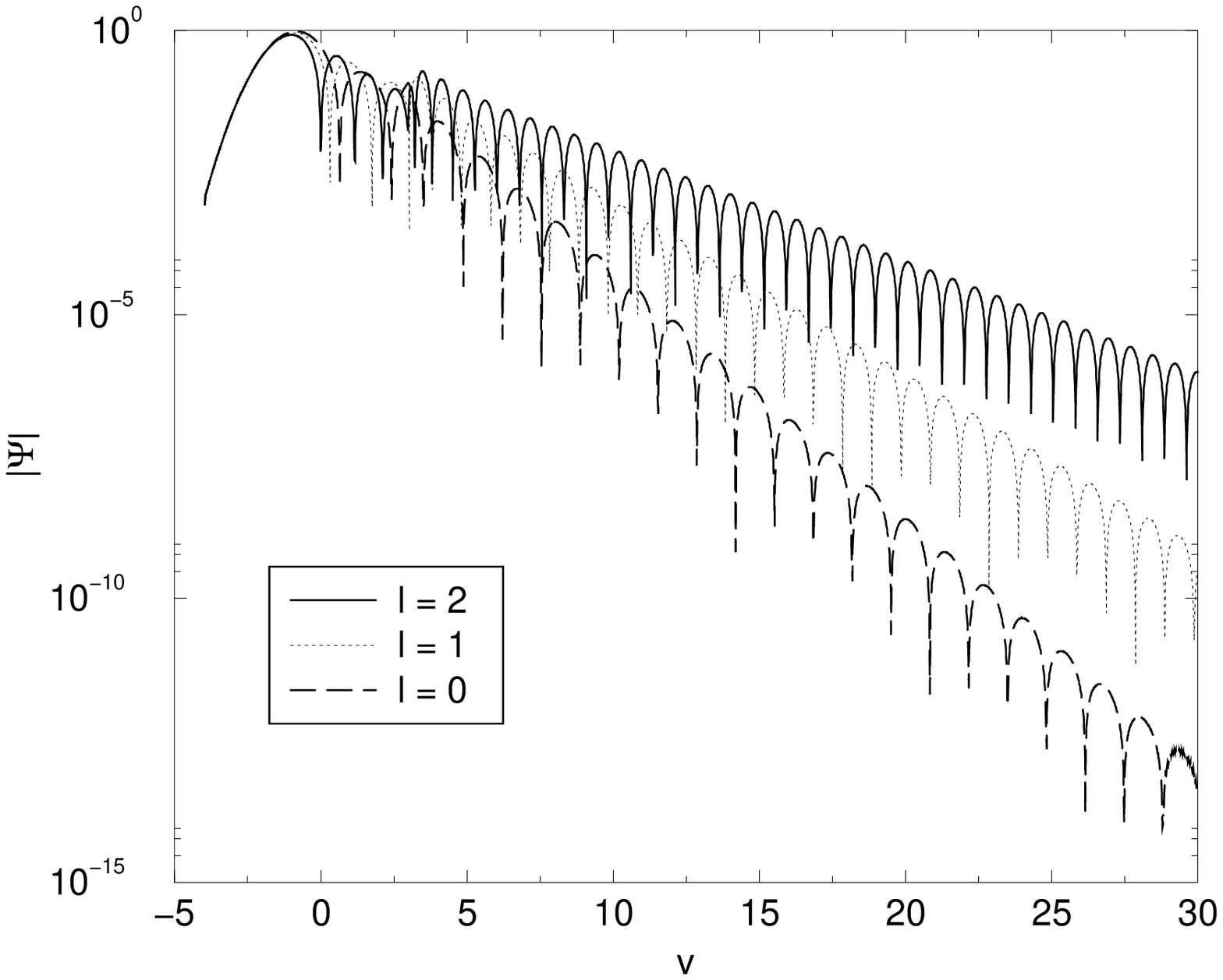}}}
\end{picture}
\parbox[t]{\textwidth}{\small Figure 4: {\it Semi--log graph of the
wave function in the $Q=0$ case for $r_+ = 0.4$, with $l=0,1,2$.}}
\end{center}

\begin{center}
\setlength{\unitlength}{0.8mm}
\begin{picture}(110,80)
\put(0,-2){\resizebox{110\unitlength}{80\unitlength}
{\includegraphics{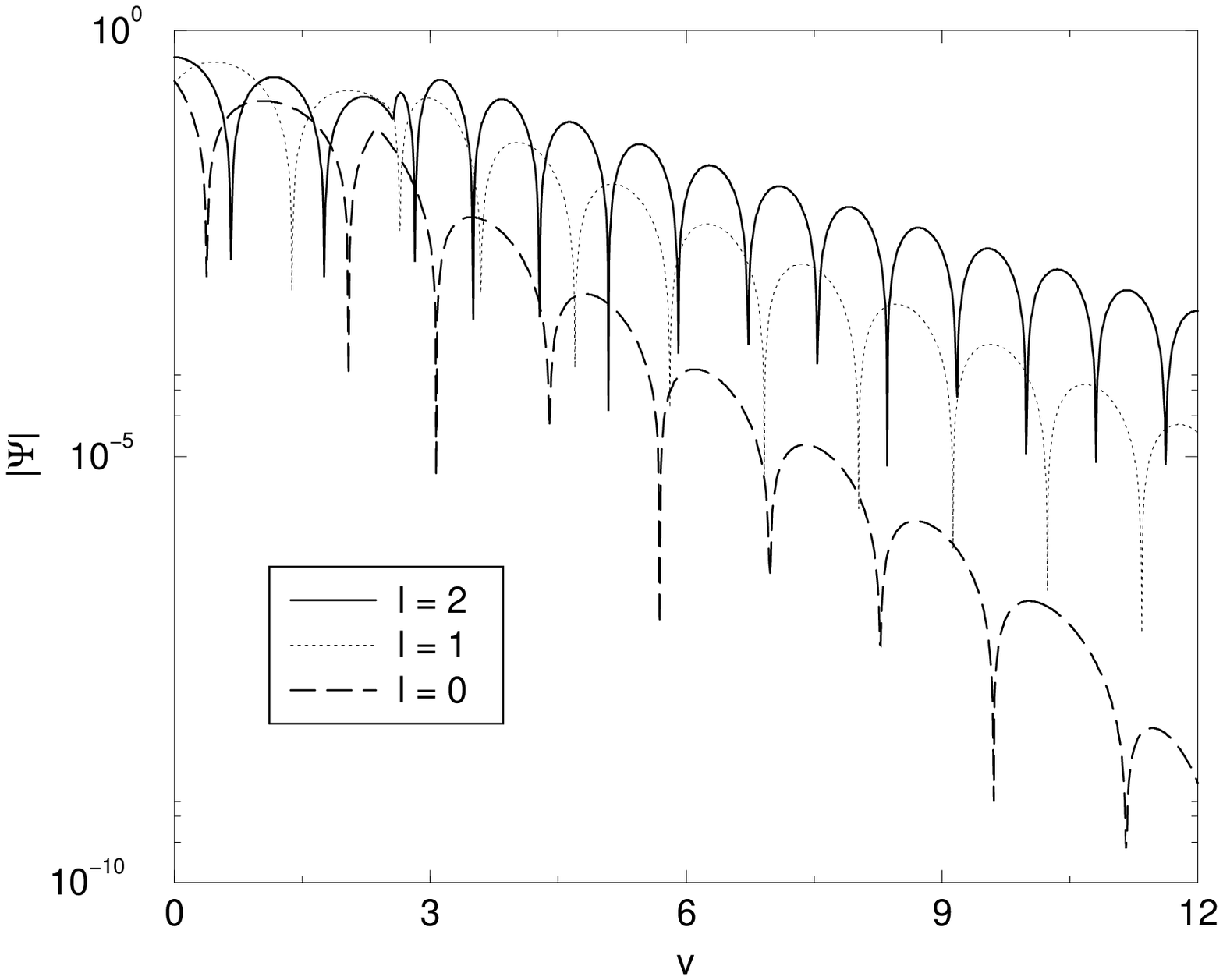}}}
\end{picture}
\parbox[t]{\textwidth}{\small Figure 5: {\it Semi--log graph of the
wave function  for $Q^2=0.1$ and $r_+ = 0.4$, with $l=0,1,2$.}}
\end{center}

\begin{center}
\setlength{\unitlength}{0.8mm}
\begin{picture}(110,80)
\put(0,-2){\resizebox{110\unitlength}{80\unitlength}
{\includegraphics{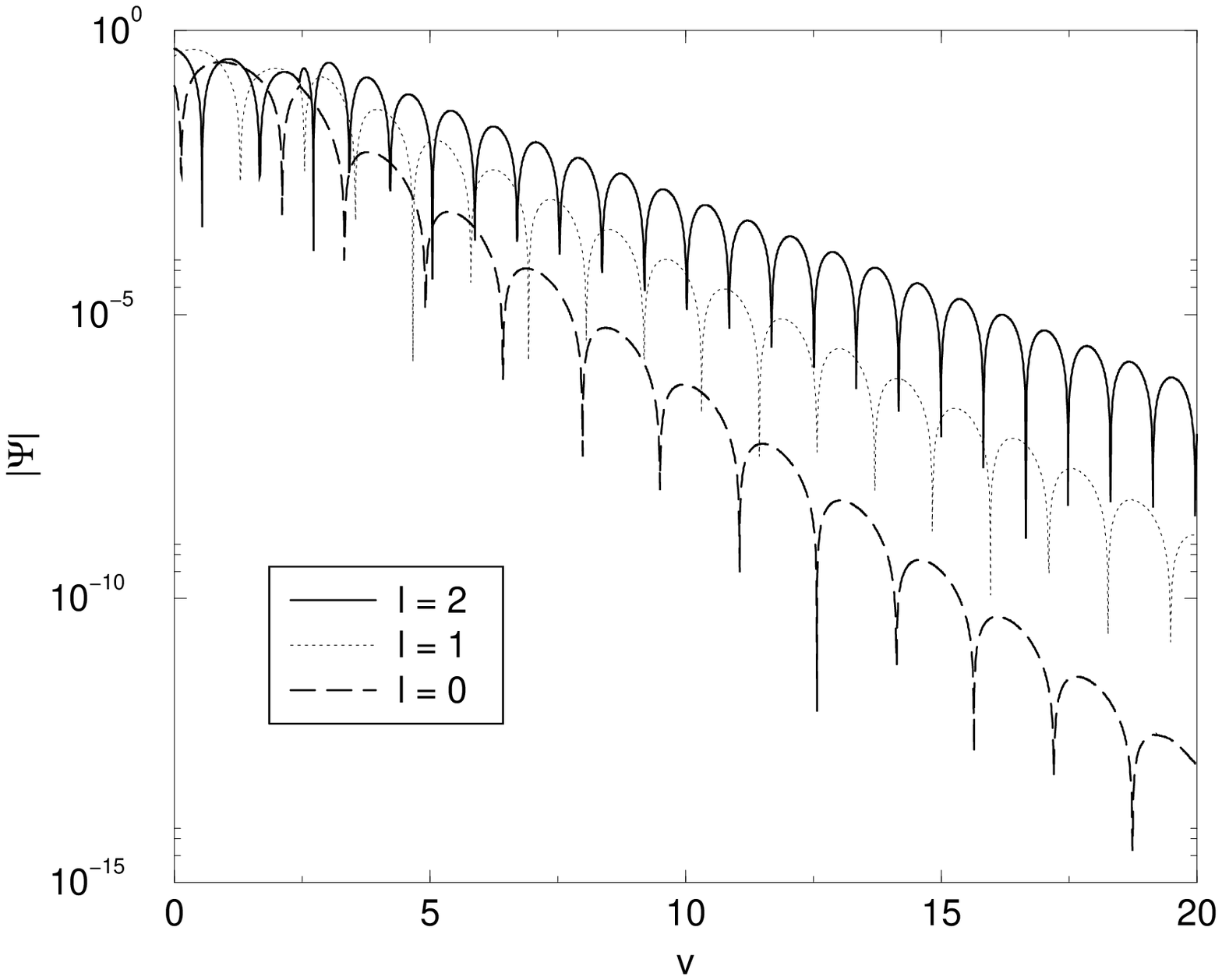}}}
\end{picture}
\parbox[t]{\textwidth}{\small Figure 6: {\it  Semi--log graph of the
wave function for $Q^2=0.125$ and $r_+ = 0.4$, with  $l=0,1,2$.}}
\end{center}

Figures 4 to 6 exhibit a striking consistent picture with those given
in \cite{Hor-Hub,Wan-Abd}. Increasing $l$, the evolution of the test
field experiences an increase of the damping time scale and a
decrease of the oscillation time scale. These figures gave us an
object--lesson on the evolution of the test field with the increase
of multipole index. It is worth noting that the behavior shown here
differs quite a lot from that of the asymptotically--flat case
\cite{Gun-Pri-Pul883}, where the perturbation settles down faster
with the increase of $l$. These differences can be used to further
support the argument that the last two processes of the wave
evolution reflect directly of the background spacetime property.

\subsection{Quasinormal modes for highly charged AdS black hole}

In our previous work \cite{Wan-Abd}, we found that there is a
numerical convergence problem when the charge $Q$ approaches the
extreme value. We claimed that the problem is related to the method
we adopted there. Using the numerical strategy proposed in
\cite{Gun-Pri-Pul883} to directly describe the wave dynamics, we
can step further on this problem.

\begin{center}
\setlength{\unitlength}{1.0mm}
\begin{picture}(150,75)(0,0)
\put(0,0){\resizebox{76\unitlength}{75\unitlength}%
{\includegraphics{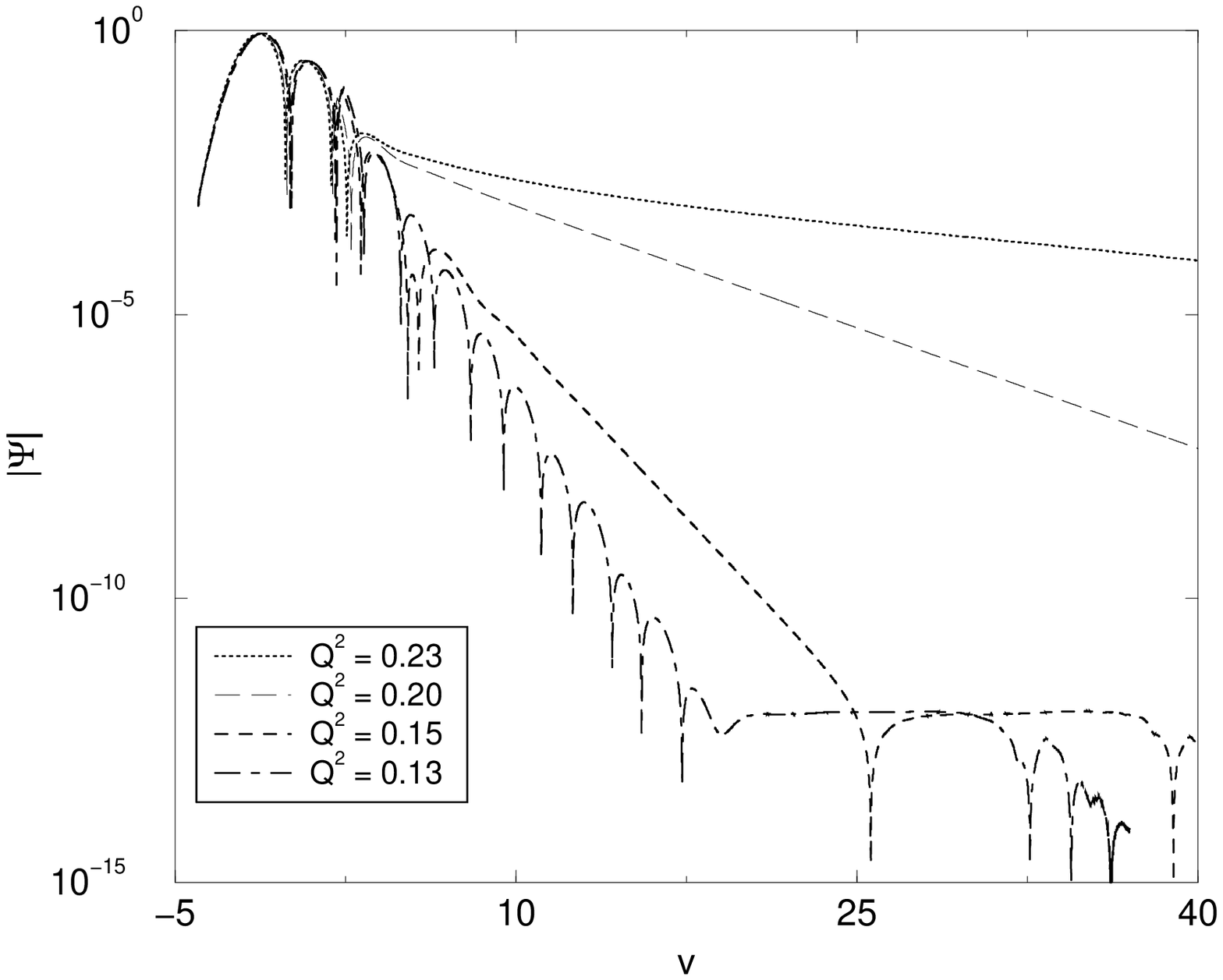}}}
\put(80,0){\resizebox{76\unitlength}{75\unitlength}%
{\includegraphics{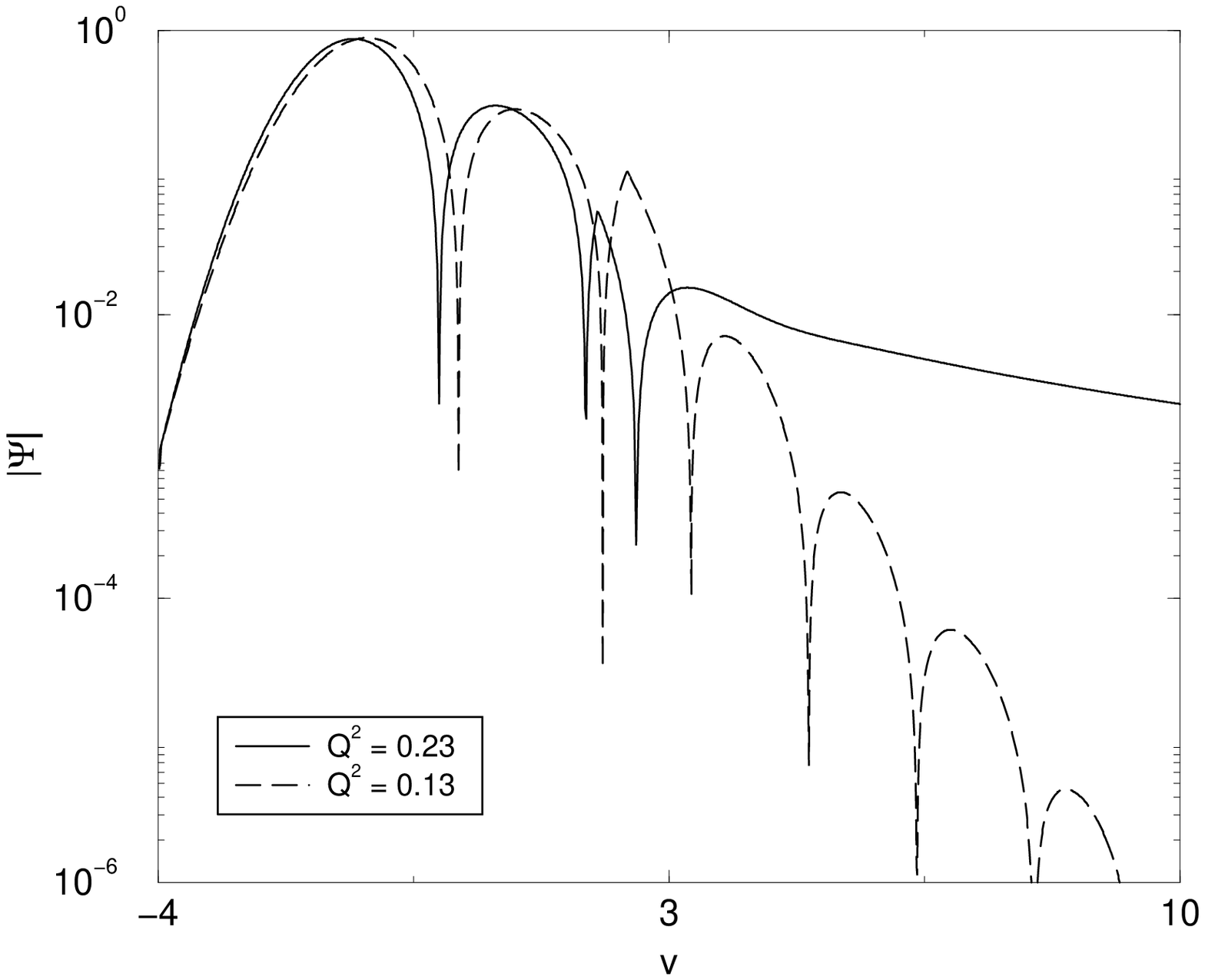}}}
\end{picture}
\parbox[t]{\textwidth}{\small Figure 7: {\it (left) Semi--log graph of
the wave function with $r_+ =0.4$, $l=0$ and several values of
$Q$. The extreme value for $Q^2$ is $0.2368$. (right) Detail of
previous plot, with two values of $Q$.}}
\end{center}

The behavior  we read from figure 7 is quite different from that
exhibited in figure 3 and discussed in subsection 3.1. Here we see
that over some critical value of $Q$, the damping time scale
increases with the increase of $Q$, corresponding to the decrease of
imaginary frequency. This means that over some critical value of
$Q$, the larger the black hole charge is, the slower for the outside
perturbation to die out.  This qualitative change in the
characteristic of the imaginary frequency as $Q$ increases agrees to
the normal--mode frequencies of RN black hole described in
\cite{And93,Kok-Sch}.

We found in the numerical calculation that the computation time
becomes bigger when $Q$ is large enough. This happens because we are
forced to take a bigger $u_{max}$, and also because the numerical
routine which calculates the function $r(r^*)$ is more time consuming
in this range of $Q$. Moreover, the precision cannot be improved for
taking smaller grid scale factor when $Q$ is close to the extreme
value. Some plateaus appear in the late--time leading us to suspect
that there may be a failure in the numerical method in this
region. In spite of this, the convergence of the numerical code
appears to be good even in this limit, as commented in appendix A.
However the question whether there is a plausible explanation for the
``wiggle'' of the imaginary frequency as the charge is large enough
and close to the extreme value is still open.

We know that the frequencies and damping times of the quasinormal
modes are entirely fixed by the black hole, and are independent of the
initial perturbation. It has been shown that there is a second order
phase transition in the extreme limit of black holes
\cite{Lou,Kab,Su-Cai-Yu}. This result has been further supported in a
recent study for the charged AdS black hole
\cite{Cha-Emp-Joh-Mye,Cha-Emp}. Whether different properties of
imaginary frequencies with the increase of the charge $Q$ reflect
different phase characteristics is a question to be answered. We are
trying to improve our numerical code to increase convergence rate and
precision. These results can be compared to those of \cite{And00}
where it has been shown that in the Kerr solution, when the rotation
parameter $a$ of the black hole approaches the critical value, the
quasinormal modes decay time gets longer. These results support the
claims of the present paper.  We will address this problem in detail
in a forthcoming paper. 

\section{Conclusions and discussions}

We have studied the wave evolution in  Reissner--Nordstr\"{o}m
Anti--de Sitter spacetimes and revealed consistent but interesting
results. The radiative tails associated with a massless scalar field
propagation on a fixed background of an AdS black hole experience an
oscillatory exponential decay at the black hole event horizon. This
result of exponential decay supports \cite{Hor-Hub} and is similar to
that on the black hole horizon in de Sitter case because of the
similar behavior of the exponentially falling off of the potential at
event horizon. When one considers the conclusions of Ching et al
\cite{Chi-Leu-Sue-You}, this late--time tail is not
surprising. Accompanied with the exponentially decay, there is a
special oscillation in the tail in linear analysis in AdS. This can be
attributed to waves bouncing off the potential at large $r$, because
the potential diverges at infinity in AdS space. The property of the
late--time behavior on the event horizon of AdS black hole can help us
better understanding the stability of Cauchy horizon in AdS spacetimes.

We have also learnt an object--lesson of the quasinormal modes on the
background  of RN AdS black hole. For small values of $Q$ the picture
we obtained is consistent with that derived in \cite{Wan-Abd}. The
larger the black hole charge is, the quicker is the approach to thermal
equilibrium. The consistent picture of the quasinormal modes depending
on the multipole index $l$ has also been illustrated. Increasing 
$l$, we obtain the effect of increasing the damping time scale and
decreasing of the oscillation time scale. 
 
When the charge in background RN AdS black hole is large enough, we get
an opposite result for the characteristic of imaginary
frequencies.  This ``wiggle'' of the imaginary frequencies as the charge
increases on the background agrees with earlier results of normal--mode
frequency study for RN black hole. Since the type of quasinormal
mode is determined by the background spacetimes and it has been found
that a second--order phase transition appears when the RN AdS black
hole becomes  extreme, we speculate that the different behavior of the
quasinormal frequencies may reflect the characteristic of two
different phases. Further study on this subject is called for.  

ACKNOWLEDGMENT:
This work was partially supported by Funda\c{c}\~{a}o de Amparo
\`{a} Pesquisa do Estado de S\~{a}o Paulo (FAPESP) and
Conselho Nacional de Desenvolvimento   Cient\'{\i}fico e
Tecnol\'{o}gico (CNPQ). B. Wang would like to acknowledge the support
given by Shanghai Science and Technology Commission.

\appendix

\section{Convergence of the Code}

As shown in expression (\ref{disc-eq}), the local truncation error
for $\Psi_N$ scales as $\epsilon^4$, assuming that the exact function
$\Psi_l$ has a Taylor expansion in the null rectangle. Although we
have little control over the global error at $u = u_{max}$, we can make a
rough estimative. The first point in the line  $u = u_{max}$, which is
$\Psi(u_{max}, v_1)$, is obtained using only points in the line $v
= v_{1}$. Since there are $\mathcal{O} (\epsilon^{-1})$ points in this
line, the global error in $\Psi_l(u_{max}, v_1)$ should scale as
$\epsilon^3$. On the other hand, since there are $\mathcal{O}
(\epsilon^{-2})$ grid points, the last point in the line  $u = u_{max}$,
which depends of all the grid points, should scale as $\epsilon^2$. 

We can improve on this estimative calculating convergence curves. In
these curves, we plot the value of $\Psi_l$ at a given point as the
grid spacing $\epsilon$ is reduced. In the figures 8 and 9 we show some curves
for three value of the charges in two points. The parameter $N$ is the
number of points in the line $u = u_0$. It is therefore inversely
proportional to  $\epsilon$.

\begin{center}
\setlength{\unitlength}{1.0mm}
\begin{picture}(110,80)
\put(5,5){\resizebox{100\unitlength}{70\unitlength}
{\includegraphics{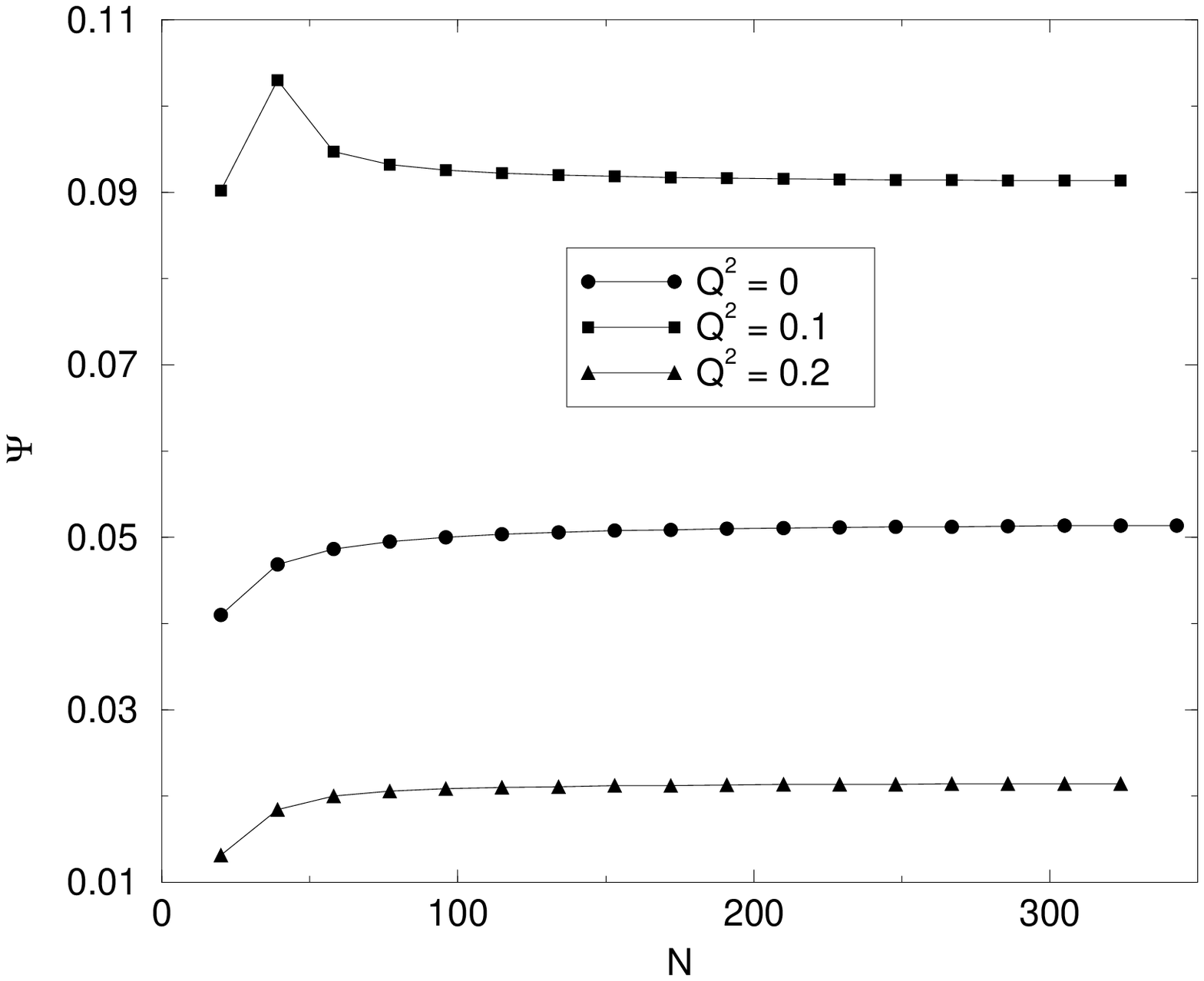}}}
\end{picture}
\parbox[t]{\textwidth}{Figure 8: {\it\small Graph of $\Psi_l$ for $v =
    2.9824$, $r_+ = 0.4$, $l = 0$ and several values  of $Q$. }}    
\vspace{5mm}
\end{center}

\begin{center}
\setlength{\unitlength}{1.0mm}
\begin{picture}(110,80)
\put(5,5){\resizebox{100\unitlength}{70\unitlength}
{\includegraphics{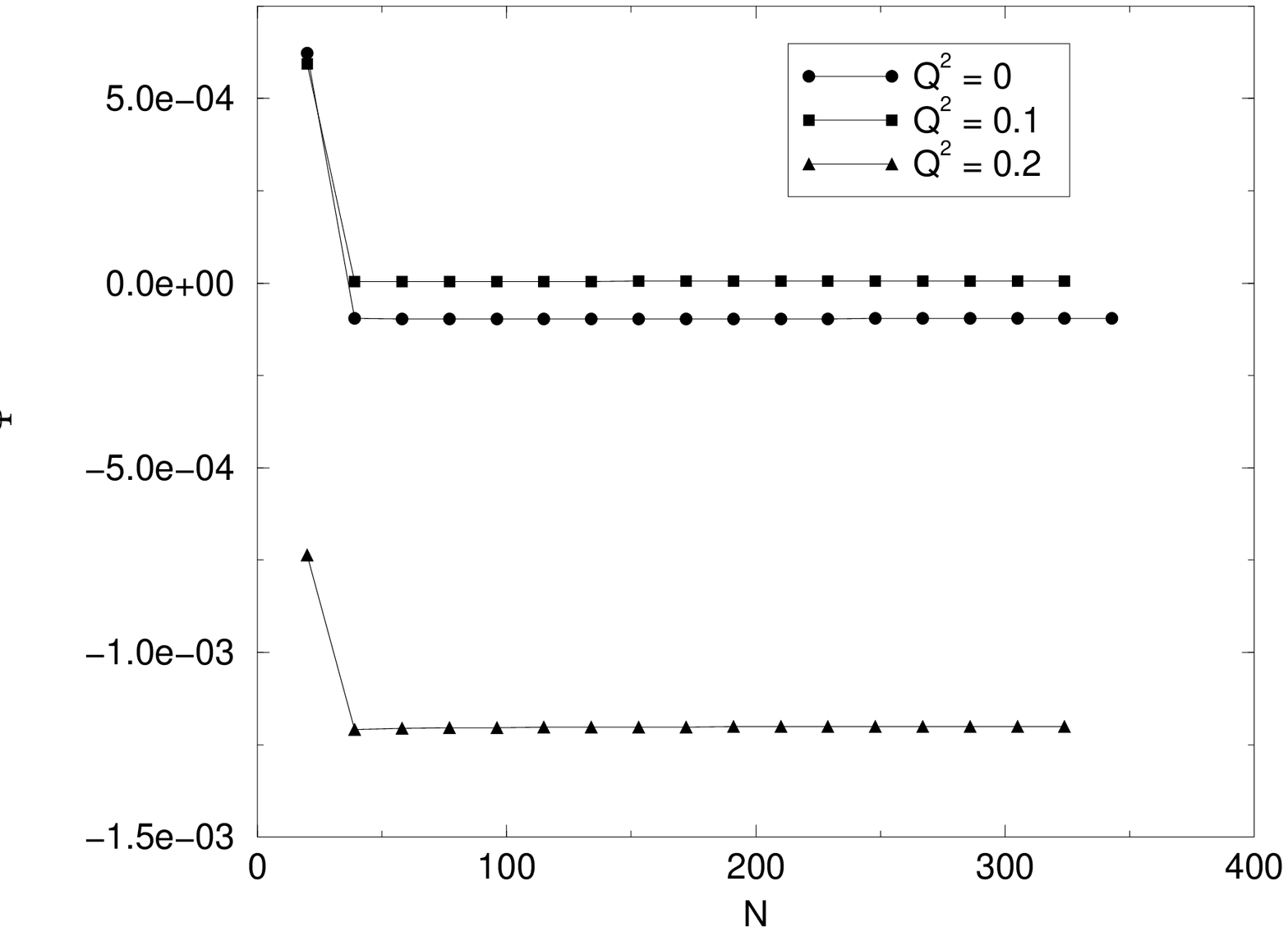}}}
\end{picture}
\parbox[t]{\textwidth}{Figure 9: {\it\small Graph of $\Psi_l$ for $v =
     9.6157$, $r_+ = 0.4$, $l = 0$ and several values  of $Q$. }}    
\vspace{5mm}
\end{center}

These diagrams clearly show numerical convergence,  and a quite
fast one. This indicates that the global error indeed tends to zero as we
decrease the grid spacing, even in the limit when $Q$ gets near the
extreme limit.

\end{document}